\documentclass[runningheads]{llncs}

\usepackage[T1]{fontenc}
\def\doi#1{\href{https://doi.org/\detokenize{#1}}{\url{https://doi.org/\detokenize{#1}}}}
\usepackage{graphicx}
\usepackage{amsmath}
\usepackage{newtxmath}
\usepackage{amsfonts}
\usepackage{multirow}
\usepackage{tabularx}
\usepackage{ulem}
\usepackage{bm}
\usepackage[dvipsnames]{xcolor}
\usepackage{threeparttable}
\usepackage{amsmath}
\usepackage[pagebackref=true,breaklinks=true,colorlinks,bookmarks=false]{hyperref}

\usepackage{booktabs}

\usepackage{color}

\usepackage{listings}
\begin{document}
\title{Spectral Adversarial MixUp for Few-Shot Unsupervised Domain Adaptation}

\author{Jiajin Zhang \inst{1}
\and Hanqing Chao \inst{1}
\and Amit Dhurandhar \inst{2}
\and Pin-Yu Chen \inst{2}
\and Ali Tajer \inst{3} 
\and Yangyang Xu \inst{4} 
\and Pingkun Yan \inst{1} \thanks{Pingkun Yan is corresponding author.}}

\authorrunning{J. Zhang et al.}
%
\institute{Department of Biomedical Engineering and Center for Biotechnology and Interdisciplinary Studies, Rensselaer Polytechnic Institute, Troy, NY, USA\\ 
\and IBM Thomas J. Watson Research Center, Yorktown Heights, NY, USA\\ 
\and Department of Electrical, Computer, and Systems Engineering, Rensselaer Polytechnic Institute, Troy, NY, USA\\
\and Department of Mathematical Sciences, Rensselaer Polytechnic Institute, Troy, NY, USA\\
}
%
%
%
%
%
\maketitle              
\begin{abstract}
Domain shift is a common problem in clinical applications, where the training images (source domain) and the test images (target domain) are under different distributions. Unsupervised Domain Adaptation (UDA) techniques have been proposed to adapt models trained in the source domain to the target domain. However, those methods require a large number of images from the target domain for model training. In this paper, we propose a novel method for Few-Shot Unsupervised Domain Adaptation (FSUDA), where only a limited number of \textit{unlabeled} target domain samples are available for training. To accomplish this challenging task, first, a spectral sensitivity map is introduced to characterize the generalization weaknesses of models in the frequency domain. We then developed a Sensitivity-guided Spectral Adversarial MixUp (SAMix) method to generate target-style images to effectively suppresses the model sensitivity, which leads to improved model generalizability in the target domain. We demonstrated the proposed method and rigorously evaluated its performance on multiple tasks using several public datasets. The source code is available at \url{https://github.com/RPIDIAL/SAMix}.

\keywords{Few-shot UDA  \and Data Augmentation \and Spectral Sensitivity.}

\end{abstract}

\section{Introduction}
A common challenge for deploying deep learning to clinical problems is the discrepancy between data distributions across different clinical sites~\cite{guan2021domain,pan2009survey,wang2022generalizing,zhang2023ucat,zhang2021task}. This discrepancy, which results from vendor or protocol differences, can cause a significant performance drop when models are deployed to a new site~\cite{chen2019synergistic,Xie_UDA,wang2020dofe}. 
To solve this problem, many Unsupervised Domain Adaptation (UDA) methods~\cite{guan2021domain} have been developed for adapting a model to a new site with only unlabeled data (target domain) by transferring the knowledge learned from the original dataset (source domain).
However, most UDA methods require sufficient target samples, which are scarce in medical imaging due to the limited accessibility to patient data. This motivates a new problem of Few-Shot Unsupervised Domain Adaptation (FSUDA), where only a few \textit{unlabeled} target samples are available for training. 

Few approaches~\cite{luo2020adversarial,wu2022style} have been proposed to tackle the problem of FSUDA.
Luo et.~al~\cite{luo2020adversarial} introduced Adversarial Style Mining (ASM), which uses a pre-trained style-transfer module to generate augmented images via an adversarial process. However, this module requires extra style images~\cite{huang2017arbitrary} for pre-training. Such images are scarce in clinical settings, and style differences across sites are subtle. This hampers the applicability of ASM to medical image analysis. SM-PPM~\cite{wu2022style} trains a style-mixing model for semantic segmentation by augmenting source domain features to a fictitious domain through random interpolation with target domain features. However, SM-PPM is specifically designed for segmentation tasks and cannot be easily adapted to other tasks.
Also, with limited target domain samples in FSUDA, the random feature interpolation is ineffective in improving the model’s generalizability.
In a different direction, numerous UDA methods have shown high performance in various tasks~\cite{tsai2018learning,vu2019advent,tang2020unsupervised,chen2022reusing}. 
However, their direct application to  FSUDA can result in severe overfitting due to the limited target domain samples~\cite{wu2022style}.
Previous studies~\cite{yang2020fda,xu2021fourier,liu2021feddg,guyader2004image} have demonstrated that transferring the amplitude spectrum of target domain images to a source domain can effectively convey image style information and diversify training dataset. To tackle the overfitting issue of existing UDA methods, we propose a novel approach called Sensitivity-guided Spectral Adversarial MixUp (SAMix) to augment training samples. 
This approach uses an adversarial mixing scheme and a spectral sensitivity map that reveals model generalizability weaknesses to generate hard-to-learn images with limited target samples efficiently.
%
%
%
%
SAMix focuses on two key aspects. \textbf{1)} \textit{Model generalizability weaknesses}: Spectral sensitivity analysis methods have been applied 
in different works~\cite{yin2019fourier} to quantify the model's spectral weaknesses to image amplitude corruptions. Zhang et al. ~\cite{zhang2022neural} demonstrated that using a spectral sensitivity map to weigh the amplitude perturbation is an effective data augmentation. However, existing sensitivity maps only use single-domain labeled data and cannot leverage target domain information. To this end, we introduce a Domain-Distance-modulated Spectral Sensitivity (DoDiSS) map to analyze the model's weaknesses in the target domain and guide our spectral augmentation.
\textbf{2)} \textit{Sample hardness}: Existing studies~\cite{wang2021augmax,luo2020adversarial} have shown that mining hard-to-learn samples in model training can enhance the efficiency of data augmentation and improve model generalization performances. Therefore, to maximize the use of the limited target domain data, we incorporate an adversarial approach into the spectral mixing process to generate the most challenging data augmentations.
This paper has three major contributions.
\textbf{1)} We propose SAMix, a novel approach for augmenting target-style samples by using an adversarial spectral mixing scheme. SAMix enables high-performance UDA methods to adapt easily to FSUDA problems.
%
\textbf{2)} 
We introduce DoDiSS to characterize a model's generalizability weaknesses in the target domain.
%
\textbf{3)} We conduct thorough empirical analyses to demonstrate the effectiveness and efficiency of SAMix as a plug-in module for various UDA methods across different tasks.

\section{Methods}

\begin{figure}[h]
    \centering
    \includegraphics[width=\columnwidth, clip=true, trim=0 2 0 2]{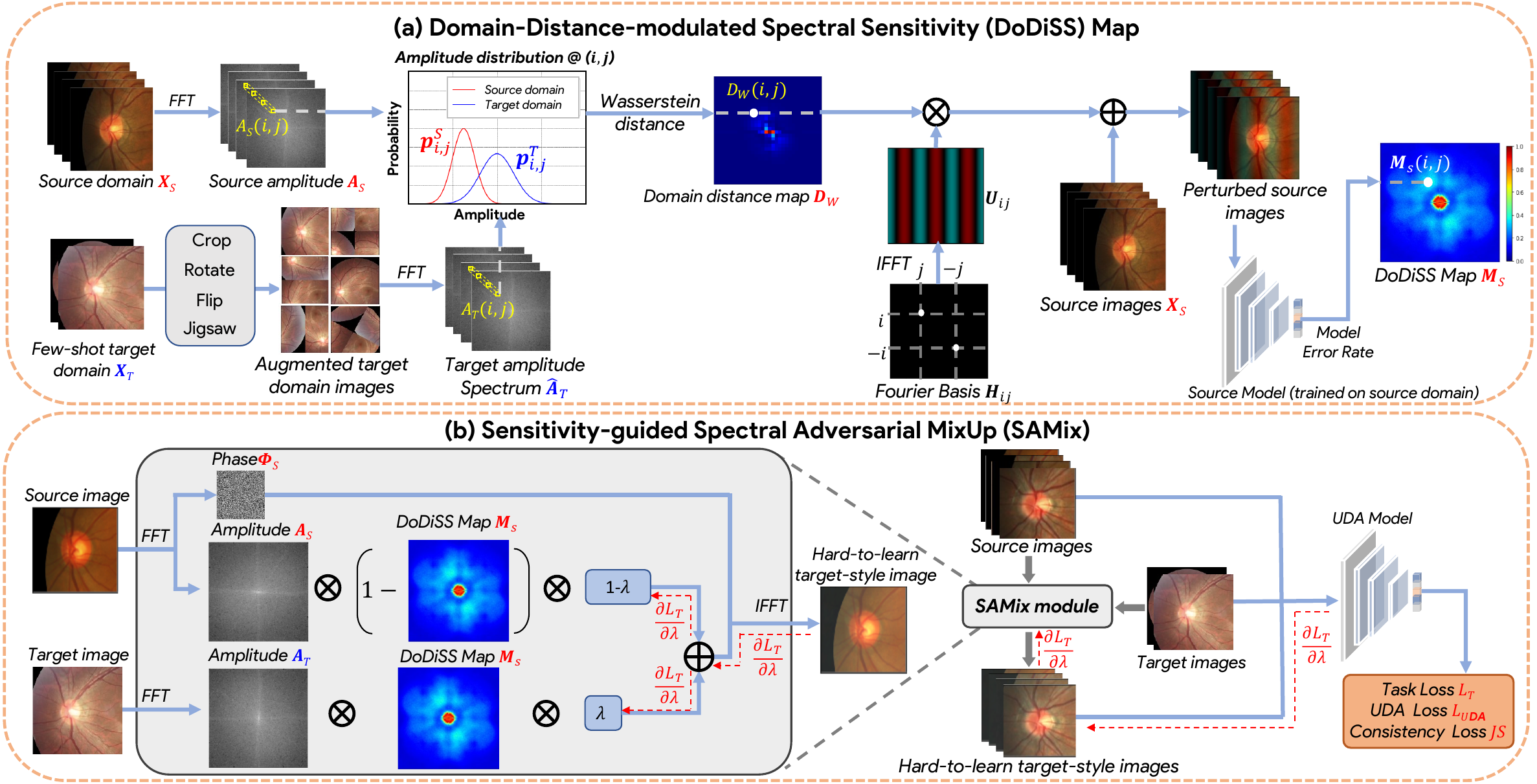}
    \caption{Illustration of the proposed framework. \textbf{(a)} DoDiSS map characterizes a model's generalizability weaknesses. \textbf{(b)} SAMix enables  UDA methods to solve FSUDA.
}
    \label{fig:method_overview}
\end{figure}

We denote the labeled source domain as $\boldsymbol{X}_S = \{(\boldsymbol{x}_n^s, \boldsymbol{y}_n^s)\}_{n=1}^N$ and the unlabeled $K$-shot target domain as $\boldsymbol{X}_T = \{\boldsymbol{x}_k^t\}_{k=1}^K$, $\boldsymbol{x}_n^s,\  \boldsymbol{x}_k^t \in \mathbb{R}^{h \times w}$. 
%
Figure~\ref{fig:method_overview} depicts the framework of our method as a plug-in module for boosting a UDA method in the FSUDA scenario. It contains two components. First, a Domain-Distance-modulated Spectral Sensitivity (DoDiSS) map is calculated to characterize a source model's weaknesses in generalizing to the target domain. Then, this sensitivity map is used for Sensitivity-guided Spectral Adversarial MixUp (SAMix) to generate target-style images for UDA models.
The details of the components are presented in the following sections.

\subsection{Domain-Distance-modulated Spectral Sensitivity (DoDiSS)}

The prior research~\cite{zhang2022neural} found that a spectral sensitivity map obtained using Fourier-based measurement of model sensitivity can effectively portray the generalizability of that model. However, the spectral sensitivity map is limited to single-domain scenarios and cannot integrate target domain information to assess model weaknesses under specific domain shifts. Thus, we introduce DoDiSS, extending the previous method by incorporating domain distance to tackle domain adaptation problems. Figure ~\ref{fig:method_overview}~\textbf{(a)} depicts the DoDiSS pipeline. It begins by computing a domain distance map for identifying the amplitude distribution difference between the source and target domains in each frequency. Subsequently, this difference map is used for weighting amplitude perturbations when calculating the DoDiSS map.

\textbf{Domain Distance Measurement.} 
To overcome the limitations of lacking target domain images, we first augment the few-shot images from the target domain with random combinations of various geometric transformations, including random cropping, rotation, flipping, and JigSaw~\cite{noroozi2016unsupervised}. These transformations keep the image intensities unchanged, preserving the target domain style information. 
The Fast Fourier Transform (FFT) is then applied to all the source images and the augmented target domain images to obtain their amplitude spectrum, denoted as $\boldsymbol{A}_S$ and $\boldsymbol{\hat{A}}_T$, respectively. We calculate the probabilistic distributions $\boldsymbol{p}^S_{i, j}$ and $\boldsymbol{p}^T_{i, j}$ of $\boldsymbol{A}_S$ and $\boldsymbol{\hat{A}}_T$ at the $(i, j)_{th}$ frequency entry, respectively.
The domain distance map at $(i,j)$ is defined as $\boldsymbol{D}_W (i,j) = W_1(\boldsymbol{p}^S_{i, j}, \boldsymbol{p}^T_{i, j})$, where $W_1$ is the 1-Wasserstein distance.



\textbf{DoDiSS Computation.} 
With the measured domain difference, we can now compute the DoDiSS map of a model.
As shown in Figure~\ref{fig:method_overview}~\textbf{(a)}, a Fourier basis is defined as a Hermitian matrix $\boldsymbol{H}_{i,j} \in \mathbb{R}^{h \times w}$ with only two non-zero elements at $(i, j)$ and $(-i, -j)$. 
A Fourier basis image $\boldsymbol{U}_{i,j}$ can be obtained by $\ell_2$-normalized Inverse Fast Fourier Transform (IFFT) of $\boldsymbol{A}_{i,j}$, \textit{i.e.}, 
$\boldsymbol{U}_{i,j} = \frac{\mathcal{IFFT}(\boldsymbol{A}_{i,j})}{||\mathcal{IFFT}(\boldsymbol{A}_{i,j})||_2}$.  
To analyze the model's generalization weakness with respect to the frequency $(i,j)$, we generate perturbed source domain images by adding the Fourier basis noise
$\boldsymbol{N}_{i,j} = r \cdot \boldsymbol{D}_W (i,j) \cdot \boldsymbol{U}_{i,j}$
to the original source domain image $\boldsymbol{x}^s$ as $\boldsymbol{x}^s + \boldsymbol{N}_{i,j}$. $\boldsymbol{D}_W (i,j)$ controls the $\ell_2$-norm of $\boldsymbol{N}_{i,j}$ and $r$ is randomly sampled to be either -1 or 1. The $\boldsymbol{N}_{i,j}$ only introduces perturbations at the frequency components $(i, j)$ to the original images. The $\boldsymbol{D}_W (i,j)$ guarantees that images are perturbed across all frequency components following the real domain shift. For RGB images, we add $\boldsymbol{N}_{i,j}$ to each channel independently following~\cite{zhang2022neural}. The sensitivity at frequency $(i,j)$ of a model $F$ trained on the source domain is defined as the prediction error rate over the whole dataset $\boldsymbol{X}_S$ as
in~\eqref{eq:sensitivity_map_dw}, where $\rm {Acc}$ denotes the prediction accuracy
%
\begin{equation}
\label{eq:sensitivity_map_dw}
\boldsymbol{M}_S (i,j) = 1 - \underset{\substack{(\boldsymbol{x}^s,\boldsymbol{y}^s) \in \boldsymbol{X}_S}}{\rm{Acc}}(F( \boldsymbol{x}^s + r \cdot \boldsymbol{D}_W (i,j) \cdot \boldsymbol{U}_{i,j} ), \boldsymbol{y}^s).
\end{equation}




\subsection{Sensitivity-guided Spectral Adversarial Mixup (SAMix)}

%
Using the DoDiSS map $\boldsymbol{M}_S$ and an adversarially learned parameter $\lambda^*$ as a weighting factor, SAMix mixes the amplitude spectrum of each source image with the spectrum of a target image. DoDiSS indicates the spectral regions where the model is sensitive to the domain difference. The parameter $\lambda^*$ mines the heard-to-learn samples to efficiently enrich the target domain samples by maximizing the task loss. 
Further, by retaining the phase of the source image, SAMix preserves the semantic meaning of the original source image in the generated target-style sample.
Specifically, as shown in Figure~\ref{fig:method_overview} \textbf{(b)}, given a source image $\boldsymbol{x}^s$ and a target image $\boldsymbol{x}^t$, we compute their amplitude and phase spectrum, denoted as $(\boldsymbol{A}^s, \boldsymbol{\Phi}^s)$ and $(\boldsymbol{A}^t, \boldsymbol{\Phi}^t)$, respectively. 
SAMix mixes the amplitude spectrum by
%
\begin{equation}
\label{eq:amplitude_mixup}
\boldsymbol{A}_{\lambda^*}^{st} =  \lambda^* \cdot \boldsymbol{M}_S \cdot \boldsymbol{A}^t + (1-\lambda^*) \cdot (1-\boldsymbol{M}_S) \cdot \boldsymbol{A}^s.
\end{equation}
%
%
%
The target-style image is reconstructed 
by $\boldsymbol{x}_{\lambda^*}^{st} = \mathcal{IFFT}(\boldsymbol{A}_{\lambda^*}^{st},\ \boldsymbol{\Phi}^{s})$.
The adversarially learned parameter $\lambda^*$ is optimized by maximizing the task loss $L_T$ using the projected gradient descent with $T$ iterations and step size of $\delta$:
\begin{equation}
\label{eq:lambda}
\lambda^* = \underset{\substack{\lambda}}{\rm{arg\ max}}\ L_T(F(\boldsymbol{x}_{\lambda}^{st};\theta), \boldsymbol{y}),\ \ \rm{s.t.}\ \ \lambda\in [0,1].
\end{equation}

In the training phase, as shown in Figure \ref{fig:method_overview}~\textbf{(b)}, the SAMix module generates a batch of augmented images, which are combined with few-shot target domain images to train the UDA model. The overall training objective is to minimize
\begin{equation}
\label{eq:loss_tot}
L_{tot}(\theta) = L_T(F(\boldsymbol{x}^{s};\theta), \boldsymbol{y}) + \mu \cdot JS(F(\boldsymbol{x}^{s};\theta), F(\boldsymbol{x}_{\lambda^*}^{st};\theta)) + L_{UDA},
\end{equation}
where $L_t$ is the supervised task loss in the source domain; $JS$ is the Jensen-Shannon divergence~\cite{zhang2022neural}, which regularizes the model predictions consistency between the source images $\boldsymbol{x}^{s}$ and their augmented versions $\boldsymbol{x}_{\lambda^*}^{st}$; $L_{UDA}$ is the training loss in the original UDA method, and $\mu$ is a weighting parameter.

\section{Experiments and Results}

We evaluated SAMix on two medical image datasets. \textbf{Fundus}~\cite{orlando2020refuge,fumero2011rim} is an optic disc and cup segmentation task. Following~\cite{wang2020dofe}, we consider images collected from different scanners as distinct domains. 
The source domain contains $400$ images of the REFUGE~\cite{orlando2020refuge} training set. 
We took 400 images from the REFUGE validation set and 159 images of RIM-One~\cite{fumero2011rim} to form the target domain 1 \& 2.
We center crop and resize the disc region to $256\times 256$ as network input. 
\textbf{Camelyon}~\cite{bandi2018detection} is a tumor tissue binary classification task across $5$ hospitals. We use the training set of Camelyon as the source domain ($302,436$ images from hospitals $1-3$) and consider the validation set ($34,904$ images from hospital $4$) and test set ($85,054$ images from the hospital $5$) as the target domains 1 and 2, respectively. All the images are resized into $256\times 256$ as network input. For all experiments, the source domain images are split into training and validation in the ratio of $4:1$. We randomly selected $K$-shot target domain images for training, while the remaining target domain images were reserved for testing.

\subsection{Implementation Details}

SAMix is evaluated as a plug-in module for four UDA models: AdaptSeg~\cite{tsai2018learning} and Advent~\cite{vu2019advent} for \textbf{Fundus}, and SRDC~\cite{tang2020unsupervised} and DALN~\cite{chen2022reusing} for \textbf{Camelyon}. For a fair comparison, we adopted the same network architecture for all the methods on each task. For \textbf{Fundus}, we use a DeepLabV2-Res101~\cite{chen2017deeplab} as the backbone with SGD optimizer for $80$ epochs. The task loss $L_t$ is the Dice loss. The initial learning rate is $0.001$, which decays by $0.1$ for every $20$ epochs. The batch size is $16$. For \textbf{Camelyon}, we use a ResNet-50~\cite{he2016deep} with SGD optimizer for $20$ epochs. $L_t$ is the binary cross-entropy loss. The initial learning rate is $0.0001$, which decays by $0.1$ every $5$ epochs. The batch size is $128$. We use the fixed weighting factor $\mu = 0.01$, iterations $T = 10$, and step size $\delta = 0.1$ in all the experiments.

\subsection{Method Effectiveness}

We demonstrate the effectiveness of SAMix by comparing it with two sets of baselines. \textit{First}, we compare the performance of UDA models with and without SAMix. \textit{Second}, we compare SAMix against other FSUDA methods~\cite{luo2020adversarial,huang2017arbitrary}.

\noindent\textbf{Fundus.} Table~\ref{tab:fundus} shows the 10-run average Dice coefficient (DSC) and Average Surface Distance (ASD) of all the methods trained with the source domain and \textbf{1-shot} target domain image. The results are evaluated in the two target domains. Compared to the model trained solely on the source domain (Source only), the performance gain achieved by UDA methods (AdaptSeg and Advent) is limited. However, incorporating SAMix as a plug-in for UDA methods (AdaptSeg+SAMix and Advent+SAMix) enhances the original UDA performance significantly ($p<0.05$). Moreover, SAMix+Advent surpasses the two FSUDA methods (ASM and SM-PPM) significantly. This improvement is primarily due to spectrally augmented target-style samples by SAMix.


\begin{table}[t!]
\centering
\caption{10-run average DSC (\%) and ASD of models on REFUGE. The best performance is in \textbf{bold} and the second best is indicated with \underline{underline}.}
\scalebox{0.72}{
\begin{threeparttable}
    \begin{tabular}{p{75pt}<{\centering}|p{30pt}<{\centering}|p{30pt}<{\centering}|p{30pt}<{\centering}|p{30pt}<{\centering}|p{30pt}<{\centering}|p{30pt}<{\centering}|p{30pt}<{\centering}|p{30pt}<{\centering}|p{30pt}<{\centering}|p{30pt}<{\centering}|p{30pt}<{\centering}|p{30pt}<{\centering}}
        \toprule
        \multirow{3}{*}{\textbf{Method}} & \multicolumn{6}{c|}{\textbf{Source Domain $\rightarrow$ Target Domain 1}} & \multicolumn{6}{c}{\textbf{Source Domain $\rightarrow$ Target Domain 2}}\\
        \cline{2-13}
        & \multicolumn{3}{c|}{DSC$(\uparrow)$} & \multicolumn{3}{c|}{ASD$(\downarrow)$} 
        & \multicolumn{3}{c|}{DSC$(\uparrow)$} & \multicolumn{3}{c}{ASD$(\downarrow)$} \\
        \cline{2-13}
        & cup & disc & avg & cup & disc & avg & cup & disc & avg & cup & disc & avg\\
        \midrule
        Source Only & 61.16$^*$ & 66.54$^*$ & 63.85$^*$ & 14.37$^*$ & 11.69$^*$ & 13.03$^*$ & 55.77$^*$ & 58.62$^*$ & 57.20$^*$ & 20.95$^*$ & 17.63$^*$ & 19.30$^*$ \\
        \midrule
        AdaptSeg & 61.45$^*$ & 66.61$^*$ & 64.03$^*$ & 13.79$^*$ & 11.47$^*$ & 12.64$^*$ & 56.67$^*$ & 60.50$^*$ & 58.59$^*$ & 20.44$^*$ & 17.97$^*$ & 19.21$^*$ \\
        Advent & 62.03$^*$ & 66.82$^*$ & 64.43$^*$ & 12.82$^*$ & 11.54$^*$ & 12.18$^*$ & 56.43$^*$ & 60.56$^*$ & 58.50$^*$ & 20.31$^*$ & 17.86$^*$ & 19.09$^*$ \\
        \midrule
        ASM & 69.18$^*$ & 71.91$^*$ & 70.05$^*$ & 8.92$^*$ & 8.35$^*$ & 8.64$^*$ & 57.79$^*$ & 61.86$^*$ & 59.83$^*$ & 19.26$^*$ & 16.94$^*$ & 18.10$^*$ \\
        SM-PPM & 74.55$^*$ & 77.62$^*$ & 76.09$^*$ & 6.09$^*$ & 5.66$^*$ & 5.88$^*$ & 59.62$^*$ & 64.17$^*$ & 61.90$^*$ & 14.52$^*$ & 12.22$^*$ & 13.37$^*$ \\
        AdaptSeg+SAMix & \textbf{76.56} & \underline{80.57} & \textbf{78.57} & \underline{4.97} & \underline{4.12} & \underline{4.55} & \underline{61.75} & \underline{66.20} & \underline{63.98} & \underline{12.75} & \underline{11.09} & \underline{11.92} \\
        Advent+SAMix & \underline{76.32} & \textbf{80.64} & \underline{78.48} & \textbf{4.90} & \textbf{3.98} & \textbf{4.44} & \textbf{62.02} & \textbf{66.35} & \textbf{64.19} & \textbf{11.97} & \textbf{10.85} & \textbf{11.41} \\
        \bottomrule
    \end{tabular}
    
    \begin{tablenotes}\footnotesize
    \item[$*$] $p<0.05$ in the one-tailed paired \textit{t}-test with Advent+SAMix.
    \end{tablenotes}
\end{threeparttable} 
}
\label{tab:fundus}
\end{table}

\begin{figure}[t]
    \centering
    \includegraphics[width=\columnwidth]{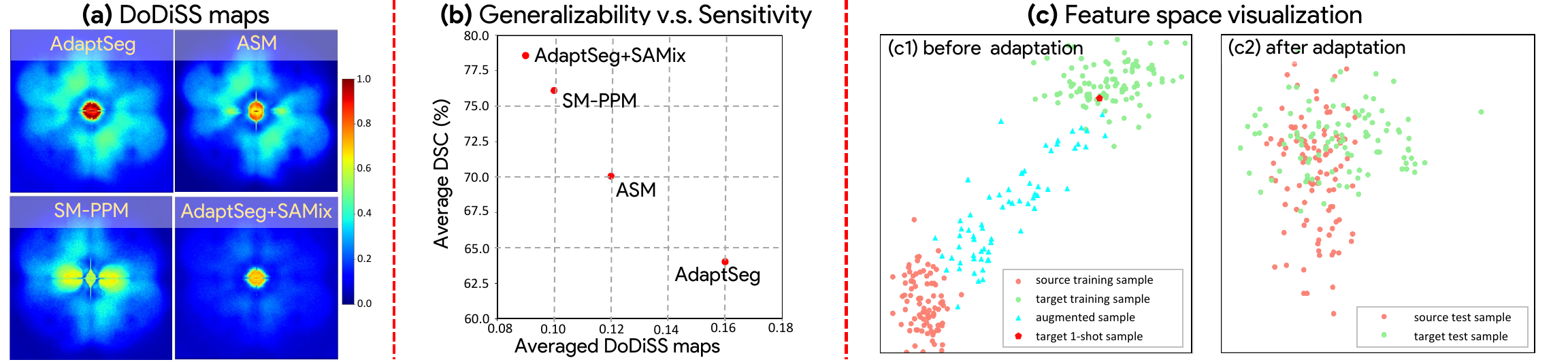}
    \caption{Method effectiveness analysis. \textbf{(a)} The DoDiSS maps visualization; \textbf{(b)} Scattering plot of model generalizability v.s. sensitivity; \textbf{(c)} Feature space visualization.}
    \label{fig:results_sense_vis}
\end{figure}

To assess the functionality of the target-aware spectral sensitivity map in measuring the model's generalization performance on the target domain, we computed the DoDiSS maps of the four models (AdaptSeg, ASM, SM-PPM, and AdaptSeg+SAMix). The results are presented in Figure~\ref{fig:results_sense_vis}\textbf{(a)}. The DoDiSS map of AdaptSeg+SAMix demonstrates a clear suppression of sensitivity, leading to improved model performance.
%
To better visualize the results, the model generalizability (average DSC) versus the averaged $\ell_1$-norm of the DoDiSS map is presented in Figure~\ref{fig:results_sense_vis} \textbf{(b)}. The figure shows a clear trend of improved model performance as the averaged DoDiSS decreases.
To assess the effectiveness of SAMix-augmented target-style images in bridging the gap of domain shift, the feature distributions of Fundus images before and after adaptation are visualized in Fig.\ref{fig:results_sense_vis} \textbf{(c)} by t-SNE~\cite{van2008visualizing}. Figure~\ref{fig:results_sense_vis}\textbf{(c1)} shows the domain shift between the source and target domain features. The augmented samples from SAMix build the connection between the two domains with only a single example image from the target domain. Please note that, except the \textbf{1-shot} sample, all the other target domain samples are used here for visualization only but never seen during training/validation. Incorporating these augmented samples in AdaptSeg merges the source and target distributions as in Figure~\ref{fig:results_sense_vis} \textbf{(c2)}.

\begin{table}[t!]
\centering
\caption{10-run average Acc (\%) and AUC (\%) of models on Camelyon. The best performance is in \textbf{bold} and the second best is indicated with \underline{underline}.}
\scalebox{0.72}{
\begin{threeparttable}
    \begin{tabular}{p{65pt}<{\centering}|p{80pt}<{\centering}|p{80pt}<{\centering}|p{80pt}<{\centering}|p{80pt}<{\centering}}
        \toprule
        \multirow{2}{*}{\textbf{Method}} & \multicolumn{2}{c|}{\textbf{Source Domain $\rightarrow$ Target Domain 1}} & \multicolumn{2}{c}{\textbf{Source Domain $\rightarrow$ Target Domain 2}}\\
        \cline{2-5}
        & Acc$(\uparrow)$ & AUC$(\uparrow)$ 
        & Acc$(\uparrow)$ & AUC$(\uparrow)$ \\
        \midrule
        Source Only & 75.42$^*$ & 71.67$^*$ & 65.55$^*$ & 60.18$^*$\\
        \midrule
        DALN & 78.63$^*$ & 74.74$^*$ & 62.57$^*$ & 56.44$^*$\\
        \midrule
        ASM & 83.66$^*$ & \underline{80.43} & 77.75$^*$ & 73.47$^*$\\
        SRDC+SAMix & \underline{84.28} & 80.05 & \underline{78.64} & \underline{74.62}\\
        DALN+SAMix & \textbf{86.41} & \textbf{82.58} & \textbf{80.84} & \textbf{75.90}\\
        \bottomrule
    \end{tabular}
    
    \begin{tablenotes}\footnotesize
    \item[$*$] $p<0.05$ in the one-tailed paired \textit{t}-test with DALN+SAMix.
    \end{tablenotes}
\end{threeparttable} 
}
\label{tab:camelyon}
\end{table}

\noindent\textbf{Camelyon.} 
The evaluation results of the 10-run average accuracy (Acc) and Area Under the receiver operating Curve (AUC) of all methods trained with \textbf{1-shot} target domain image are presented in Table~\ref{tab:camelyon}. The clustering-based SRDC is not included in the table, as the model crashed in this few-shot scenario. Also, the SM-PPM is not included because it is specifically designed for segmentation tasks. The results suggest that combining SAMix with UDA not only enhances the original UDA performance but also significantly outperforms other FSUDA methods.

\subsection{Data Efficiency}

\begin{figure}[t]
    \centering
    \includegraphics[width=\columnwidth]{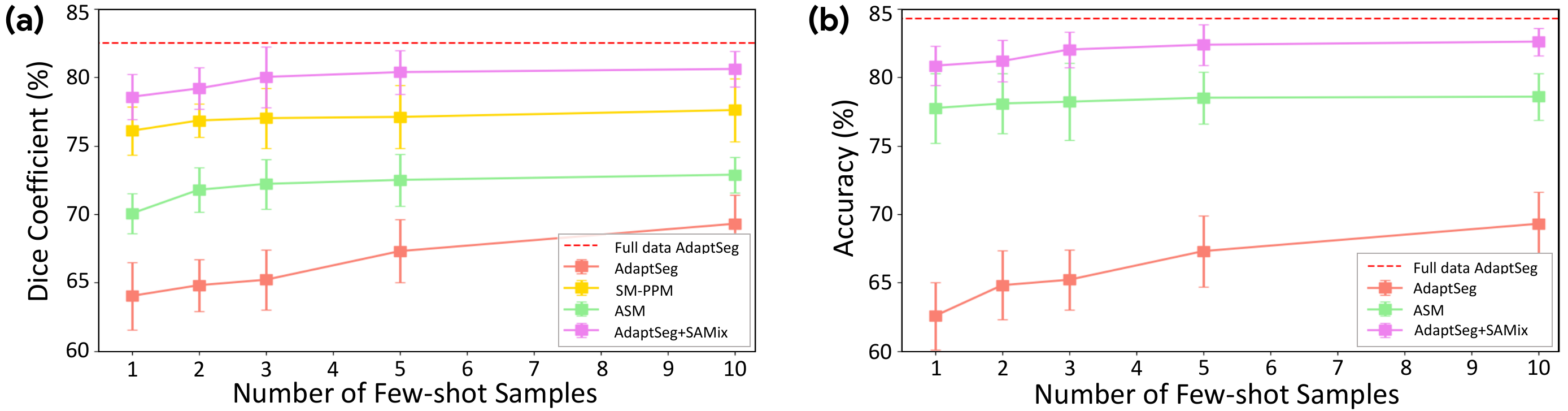}
    \caption{Data efficiency of FSUDA methods on \textbf{(a)} \textbf{Fundus} and \textbf{(b)} \textbf{Camelyon}.}
    \label{fig:multi_shot}
\end{figure}

\begin{figure}[t]
    \centering
    \includegraphics[width=\columnwidth, clip=true, trim=0 0 0 0]{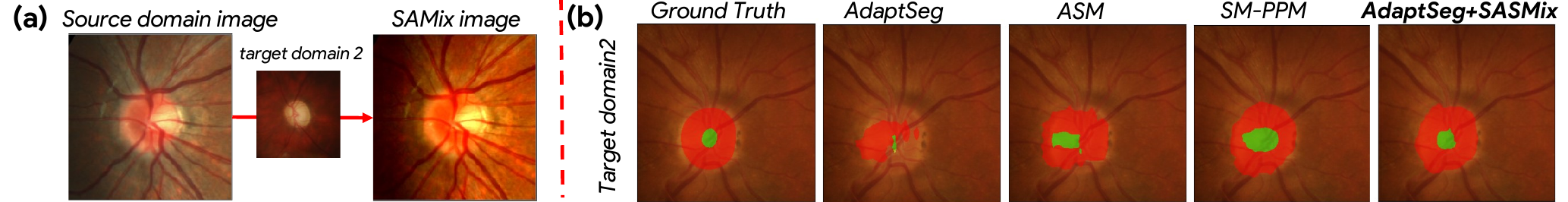}
    \caption{\textbf{(a)} SAMix generated samples. \textbf{(b)} Case study of the Fundus segmentation.}
    \label{fig:results_seg_vis}
\end{figure}

As the availability of target domain images is limited, data efficiency plays a crucial role in determining the data augmentation performance. Therefore, we evaluated the model's performance with varying numbers of target domain images in the training process. Figure~\ref{fig:multi_shot} \textbf{(a)} and \textbf{(b)} illustrate the domain adaptation results on \textbf{Fundus} and \textbf{Camelyon} (both in target domain 1), respectively. Our method consistently outperforms other baselines with just a 1-shot target image for training.
Furthermore, we qualitatively showcase the data efficiency of SAMix. Figure~\ref{fig:results_seg_vis} \textbf{(a)} displays the generated image of SAMix given the target domain image. While maintaining the retinal structure of the source image, the augmented images exhibit a more similar style to the target image, indicating SAMix can effectively transfer the target domain style.
Figure~\ref{fig:results_seg_vis} \textbf{(b)} shows an example case of the segmented results. Compared with other baselines, the SAMix segmentation presents much less prediction error, especially in the cup region.

\begin{figure}[h!]
    \centering
    \includegraphics[width=\columnwidth]{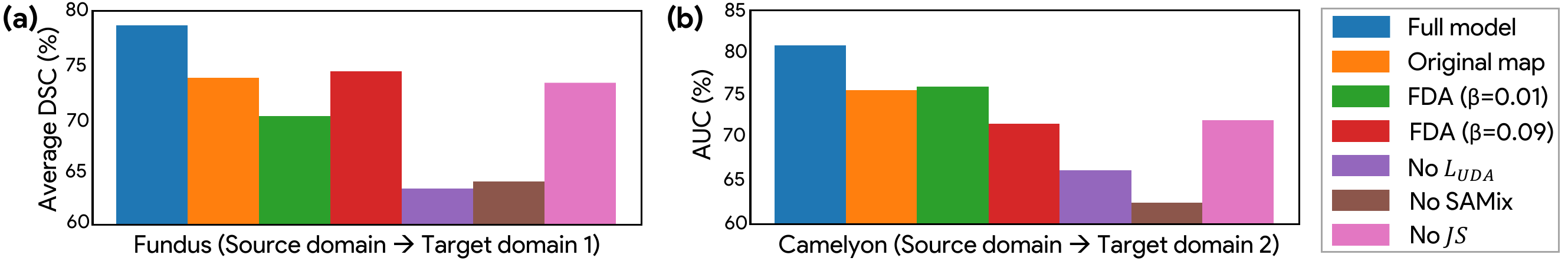}
    \caption{Ablation study. \textbf{(a)} Average DSC on \textbf{Fundus}. \textbf{(b)} AUC on \textbf{Camelyon}.}
    \label{fig:ablation}
\end{figure}

\subsection{Ablation Study}

To assess the efficacy of the components in SAMix, we conducted an ablation study with AdaptSeg+SAMix and DALN+SAMix (Full model) on Fundus and Camelyon datasets. This was done by \textbf{1)} replacing our proposed DoDiSS map with the original one in~\cite{zhang2022neural} (Original map); \textbf{2)} replacing the SAMix module with the random spectral swapping (FDA, $\beta$ = 0.01, 0.09) in~\cite{yang2020fda}; \textbf{3)} removing the three major components (No $L_{UDA}$, No SAMix, No $JS$) in a leave-one-out manner. Figure~\ref{fig:ablation} suggests that, compared with the Full model, the model performance degrades when the proposed components are either removed or replaced by previous methods, which indicates the efficacy of the SAMix components.


\section{Discussion and Conclusion}

This paper introduces a novel approach, Sensitivity-guided Spectral Adversarial MixUp (SAMix), which utilizes an adversarial mixing scheme and a spectral sensitivity map to generate target-style samples effectively. The proposed method facilitates the adaptation of existing UDA methods in the few-shot scenario. Thorough empirical analyses demonstrate the effectiveness and efficiency of SAMix as a plug-in module for various UDA methods across multiple tasks.

\section*{Acknowledgments}

This research was partially supported by the National Science Foundation (NSF) under the CAREER award OAC 2046708, the National Institutes of Health (NIH) under award R21EB028001, and the Rensselaer-IBM AI Research Collaboration of the IBM AI Horizons Network.

%
%
%
%
\newpage
\bibliographystyle{splncs04}
\bibliography{refs}

\begin{thebibliography}{10}
\providecommand{\url}[1]{\texttt{#1}}
\providecommand{\urlprefix}{URL }
\providecommand{\doi}[1]{https://doi.org/#1}

\bibitem{bandi2018detection}
Bandi, P., Geessink, O., Manson, Q., Van~Dijk, M., Balkenhol, M., Hermsen, M.,
  Bejnordi, B.E., Lee, B., Paeng, K., Zhong, A., et~al.: From detection of
  individual metastases to classification of lymph node status at the patient
  level: the camelyon17 challenge. IEEE Transactions on Medical Imaging  (2018)

\bibitem{chen2019synergistic}
Chen, C., Dou, Q., Chen, H., Qin, J., Heng, P.A.: Synergistic image and feature
  adaptation: Towards cross-modality domain adaptation for medical image
  segmentation. In: Proceedings of the AAAI conference on artificial
  intelligence. vol.~33, pp. 865--872 (2019)

\bibitem{chen2017deeplab}
Chen, L.C., Papandreou, G., Kokkinos, I., Murphy, K., Yuille, A.L.: Deeplab:
  Semantic image segmentation with deep convolutional nets, atrous convolution,
  and fully connected crfs. IEEE transactions on pattern analysis and machine
  intelligence  \textbf{40}(4),  834--848 (2017)

\bibitem{chen2022reusing}
Chen, L., Chen, H., Wei, Z., Jin, X., Tan, X., Jin, Y., Chen, E.: Reusing the
  task-specific classifier as a discriminator: Discriminator-free adversarial
  domain adaptation. In: Proceedings of the IEEE/CVF Conference on Computer
  Vision and Pattern Recognition. pp. 7181--7190 (2022)

\bibitem{fumero2011rim}
Fumero, F., Alay{\'o}n, S., Sanchez, J.L., Sigut, J., Gonzalez-Hernandez, M.:
  Rim-one: An open retinal image database for optic nerve evaluation. In: 2011
  24th international symposium on computer-based medical systems (CBMS).
  pp.~1--6. IEEE (2011)

\bibitem{guan2021domain}
Guan, H., Liu, M.: Domain adaptation for medical image analysis: a survey. IEEE
  Transactions on Biomedical Engineering  \textbf{69}(3),  1173--1185 (2021)

\bibitem{guyader2004image}
Guyader, N., Chauvin, A., Peyrin, C., H{\'e}rault, J., Marendaz, C.: Image
  phase or amplitude? rapid scene categorization is an amplitude-based process.
  Comptes Rendus Biologies  \textbf{327}(4),  313--318 (2004)

\bibitem{he2016deep}
He, K., Zhang, X., Ren, S., Sun, J.: Deep residual learning for image
  recognition. In: Proceedings of the IEEE conference on computer vision and
  pattern recognition. pp. 770--778 (2016)

\bibitem{huang2017arbitrary}
Huang, X., Belongie, S.: Arbitrary style transfer in real-time with adaptive
  instance normalization. In: Proceedings of the IEEE international conference
  on computer vision. pp. 1501--1510 (2017)

\bibitem{liu2021feddg}
Liu, Q., Chen, C., Qin, J., Dou, Q., Heng, P.A.: Feddg: Federated domain
  generalization on medical image segmentation via episodic learning in
  continuous frequency space. In: Proceedings of the IEEE/CVF Conference on
  Computer Vision and Pattern Recognition. pp. 1013--1023 (2021)

\bibitem{luo2020adversarial}
Luo, Y., Liu, P., Guan, T., Yu, J., Yang, Y.: Adversarial style mining for
  one-shot unsupervised domain adaptation. Advances in Neural Information
  Processing Systems  \textbf{33},  20612--20623 (2020)

\bibitem{van2008visualizing}
Van~der Maaten, L., Hinton, G.: Visualizing data using t-sne. Journal of
  machine learning research  \textbf{9}(11) (2008)

\bibitem{noroozi2016unsupervised}
Noroozi, M., Favaro, P.: Unsupervised learning of visual representations by
  solving jigsaw puzzles. In: Computer Vision--ECCV 2016: 14th European
  Conference, Amsterdam, The Netherlands, October 11-14, 2016, Proceedings,
  Part VI. pp. 69--84. Springer (2016)

\bibitem{orlando2020refuge}
Orlando, J.I., Fu, H., Breda, J.B., Van~Keer, K., Bathula, D.R., Diaz-Pinto,
  A., Fang, R., Heng, P.A., Kim, J., Lee, J., et~al.: Refuge challenge: A
  unified framework for evaluating automated methods for glaucoma assessment
  from fundus photographs. Medical image analysis  \textbf{59},  101570 (2020)

\bibitem{pan2009survey}
Pan, S.J., Yang, Q.: A survey on transfer learning. IEEE Transactions on
  knowledge and data engineering  \textbf{22}(10),  1345--1359 (2009)

\bibitem{tang2020unsupervised}
Tang, H., Chen, K., Jia, K.: Unsupervised domain adaptation via structurally
  regularized deep clustering. In: Proceedings of the IEEE/CVF conference on
  computer vision and pattern recognition. pp. 8725--8735 (2020)

\bibitem{tsai2018learning}
Tsai, Y.H., Hung, W.C., Schulter, S., Sohn, K., Yang, M.H., Chandraker, M.:
  Learning to adapt structured output space for semantic segmentation. In:
  Proceedings of the IEEE conference on computer vision and pattern
  recognition. pp. 7472--7481 (2018)

\bibitem{vu2019advent}
Vu, T.H., Jain, H., Bucher, M., Cord, M., P{\'e}rez, P.: Advent: Adversarial
  entropy minimization for domain adaptation in semantic segmentation. In:
  Proceedings of the IEEE/CVF Conference on Computer Vision and Pattern
  Recognition. pp. 2517--2526 (2019)

\bibitem{wang2021augmax}
Wang, H., Xiao, C., Kossaifi, J., Yu, Z., Anandkumar, A., Wang, Z.: Augmax:
  Adversarial composition of random augmentations for robust training. Advances
  in neural information processing systems  \textbf{34},  237--250 (2021)

\bibitem{wang2022generalizing}
Wang, J., Lan, C., Liu, C., Ouyang, Y., Qin, T., Lu, W., Chen, Y., Zeng, W.,
  Yu, P.: Generalizing to unseen domains: A survey on domain generalization.
  IEEE Transactions on Knowledge and Data Engineering  (2022)

\bibitem{wang2020dofe}
Wang, S., Yu, L., Li, K., Yang, X., Fu, C.W., Heng, P.A.: Dofe: Domain-oriented
  feature embedding for generalizable fundus image segmentation on unseen
  datasets. IEEE Transactions on Medical Imaging  (2020)

\bibitem{wu2022style}
Wu, X., Wu, Z., Lu, Y., Ju, L., Wang, S.: Style mixing and patchwise
  prototypical matching for one-shot unsupervised domain adaptive semantic
  segmentation. In: Proceedings of the AAAI Conference on Artificial
  Intelligence. vol.~36, pp. 2740--2749 (2022)

\bibitem{Xie_UDA}
Xie, Q., Li, Y., He, N., Ning, M., Ma, K., Wang, G., Lian, Y., Zheng, Y.:
  Unsupervised domain adaptation for medical image segmentation by
  disentanglement learning and self-training. IEEE Transactions on Medical
  Imaging pp.~1--1 (2022). \doi{10.1109/TMI.2022.3192303}

\bibitem{xu2021fourier}
Xu, Q., Zhang, R., Zhang, Y., Wang, Y., Tian, Q.: A fourier-based framework for
  domain generalization. In: Proceedings of the IEEE/CVF Conference on Computer
  Vision and Pattern Recognition. pp. 14383--14392 (2021)

\bibitem{yang2020fda}
Yang, Y., Soatto, S.: Fda: Fourier domain adaptation for semantic segmentation.
  In: Proceedings of the IEEE/CVF Conference on Computer Vision and Pattern
  Recognition. pp. 4085--4095 (2020)

\bibitem{yin2019fourier}
Yin, D., Gontijo~Lopes, R., Shlens, J., Cubuk, E.D., Gilmer, J.: A fourier
  perspective on model robustness in computer vision. Advances in Neural
  Information Processing Systems  \textbf{32} (2019)

\bibitem{zhang2022neural}
Zhang, J., Chao, H., Dhurandhar, A., Chen, P.Y., Tajer, A., Xu, Y., Yan, P.:
  When neural networks fail to generalize? a model sensitivity perspective. In:
  Proceedings of the AAAI Conference on Artificial Intelligence (2023)

\bibitem{zhang2021task}
Zhang, J., Chao, H., Xu, X., Niu, C., Wang, G., Yan, P.: Task-oriented low-dose
  ct image denoising. In: Medical Image Computing and Computer Assisted
  Intervention--MICCAI 2021: 24th International Conference, Strasbourg, France,
  September 27--October 1, 2021, Proceedings, Part VI 24. pp. 441--450.
  Springer (2021)

\bibitem{zhang2023ucat}
Zhang, J., Chao, H., Yan, P.: Toward adversarial robustness in unlabeled target
  domains. IEEE Transactions on Image Processing  \textbf{32},  1272--1284
  (2023). \doi{10.1109/TIP.2023.3242141}

\end{thebibliography}
\end{document}